\documentclass[12pt]{iopart}

\usepackage[dvipdfm]{graphicx}
\usepackage{iopams} 
\usepackage{multirow} 
\newcommand{\Part}[3]{ \frac{ \partial^{#3} #1 }{ \partial #2^{#3} }}
 
\begin{document}
\title[Random-field $p$-spin glass model on regular random graphs]{Random-field $p$-spin glass model on regular random graphs}

\author{Yoshiki Matsuda$^{1,4}$, Hidetoshi Nishimori$^{1}$, Lenka Zdeborov\'a$^{2}$ and Florent Krzakala$^{3}$}
\address{$^{1}$Department of Physics, Tokyo Institute of Technology,
Oh-okayama, Meguro-ku, Tokyo 152-8551, Japan}
\address{$^{2}$ Institue de Physique Th\'eorique, IPhT, CEA Saclay, and URA 2306, CNRS, 91191 Gif-sur-Yvette cedex, France}
\address{$^{3}$
CNRS and ESPCI ParisTech, 10 rue Vauquelin, UMR 7083 Gulliver, Paris 75005 France
}
\ead{$^{4}$matsuda@stat.phys.titech.ac.jp}

\begin{abstract}
We investigate in detail the phase diagrams of the $p$-body $\pm J$ Ising model with and without random fields on random graphs with fixed connectivity. 
One of our most interesting findings is that a thermodynamic spin glass phase is present in the three-body purely ferromagnetic model in random fields, unlike for the canonical two-body interaction random-field Ising model. 
We also discuss the location of the phase boundary between the paramagnetic and spin glass phases that does not depend on the change of the ferromagnetic bias. 
 This behavior is explained by a gauge transformation, which shows that gauge-invariant properties generically do not depend on the strength of the ferromagnetic bias for the $\pm J$ Ising model on regular random graphs.
\end{abstract}


\maketitle

\section{Introduction}
Two typical examples of disordered system in statistical physics are the spin glass and the random-field Ising model. The basic model of the former is the Edwards-Anderson model~\cite{Edwards:75} with quenched randomness in interactions. The Sherrington-Kirkpatrick model~\cite{Sherrington:75} is the infinite-range (fully-connected) version of the Edwards-Anderson model and is considered to constitute the mean-field paradigm of the theory of spin glasses. The exact solution of the Sherrington-Kirkpatrick model is now established and is known to possess quite unusual properties represented by the symmetry breaking in the abstract space of replicas~\cite{M'ezard:87, Fischer:91, Young:98, Nishimori:01, M'ezard:09}. The spin glass theory has also been applied to a variety of problem in other disciplines including information theory~\cite{Nishimori:01,M'ezard:09}.

The Edwards-Anderson model on the random graph is closer to real physical systems than the Sherrington-Kirkpatrick model because the former has a finite connectivity whereas the latter is fully connected although both are of mean-field nature. Random systems on the tree-like lattice can be analyzed by the cavity method~\cite{M'ezard:87}. Stability of the replica-symmetric (RS) state as formulated by de Almeida and Thouless for the Sherrington-Kirkpatrick model~\cite{Almeida:78} can also be analyzed on the Bethe lattice~\cite{Thouless:86, Rivoire:04, Martin:05}, and the effects of replica-symmetry breaking (RSB) have been studied by a generalization of the cavity method~\cite{M'ezard:01}. The problem of spin glasses on the finitely-connected sparse graphs in general has also been a useful platform to formulate random optimization problems
~\cite{M'ezard:09}.

As for the random-field Ising model (RFIM), the infinite-range version has been solved by Aharony~\cite{Aharony:78}. This model has ferromagnetic and paramagnetic phases but no spin-glass phase. The boundary between the two phases represents a second-order phase transition for small values of the field strength and a first-order transition for large values if the distribution of random fields is bimodal. The RFIM on a finite connectivity graph also shows a first-order phase transition for larger values of connectivity than three~\cite{Bruinsma:84, Wohlman:84}. For the three-dimensional RFIM the replica field theory has suggested the existence of a replica-symmetry-broken (RSB) phase (spin-glass phase) in the vicinity of the phase boundary~\cite{Young:98, Almeida:87, M'ezard:90, M'ezard:92, M'ezard:94, DeDominicis:95, Br'ezin:98, Br'ezin:01, Pastor:02}. However, nonexistence of the spin-glass phase in the RFIM with two-body interactions has been proved recently for all lattices and field distributions~\cite{Krzakala:10}, and the problem of the existence of a spin-glass phase in the RFIM has thus been solved negatively as long as the interactions are two-body and purely ferromagnetic.

It is therefore interesting to study in more detail the interplay of randomness in interactions and fields on the random graph, in particular in the presence of many-body interactions. Our numerical results show that the phase boundary between the spin glass and paramagnetic phases does not change if we vary the ferromagnetic bias $\rho$, the probability that an interaction on a bond is ferromagnetic in the $\pm J$ model. This has been known for the $\pm J$ Ising model without field on the regular random graph as a horizontal paramagnetic-spin glass boundary on the $\rho$-$T$ phase diagram. Our result generalizes this knowledge to phase diagrams on the $H$-$T$ plane and the $\rho$-$H$ plane, where $H$ is the magnitude (strength) of symmetrically distributed random fields. 
This $\rho$-independence can be understood by a simple gauge transformation. 
We also show that the RFIM has a thermodynamic spin-glass phase if it has three-body even purely ferromagnetic interactions. Thus the proof for the absence of a spin glass phase in the RFIM~\cite{Krzakala:10} cannot be generalized to more than two-body interactions. This result reinforces previous findings on the glassy behavior of ferromagnetic systems~\cite{Bouchaud:94,Marinari:94,Franz:95,Franz:01}.

\section{Formulation}

In this section, we formulate the $\pm J$ Ising model in random fields on the regular random graph. We review the cavity method, which enables us to study infinite-size systems numerically and analytically.

\subsection{Cavity method}
\begin{figure}[htb]
\begin{center}
   \input{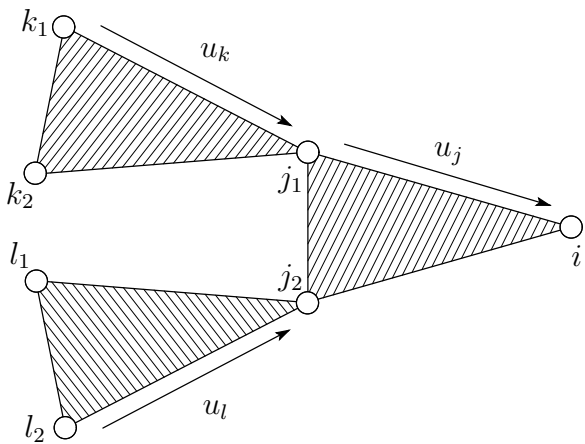}
 \caption{The local structure of the three-body ($p=3$) Ising model on the tree with the connectivity $c=2$.}
 \label{fig:lattice}
\end{center}
\end{figure}

We consider the $\pm J$ Ising model in random fields with $p$-body interactions on the random lattice having connectivity $c$ (figure~\ref{fig:lattice}). The Hamiltonian is
\begin{equation}
\mathcal{H} =  - \sum\limits_{\left\langle {i,j_1,\cdots,j_{p-1}} \right\rangle } {J_{ij_1 \cdots j_{p-1}} S_i S_{j_1} \cdots S_{j_{p-1}} }  - H \sum\limits_{i=1}^{N} {\sigma _i S_i } ,
\label{rbrfim}
\end{equation}
where $\sigma _i$ is randomly distributed as 
\begin{equation}
P_\sigma\left( {\sigma _i } \right) = \frac{1}{2}\delta \left( {\sigma _i  - 1} \right) + \frac{1}{2}\delta \left( {\sigma _i  + 1} \right)
\end{equation}
and $J_{ij_1 \cdots j_{p-1}}$ follows
\begin{equation}
P_J\left( J_{ij_1 \cdots j_{p-1}} \right) = \rho \delta \left( J_{ij_1 \cdots j_{p-1}}  - 1 \right) + \left( 1-\rho \right) \delta \left( J_{ij_1 \cdots j_{p-1}} + 1 \right),
\end{equation}
where $1/2\le \rho\le1$.

Our definition of the regular random graph is a randomly connected graph with fixed connectivity $c$ \cite{M'ezard:01, Obuchi:09, Matsuda:10}. In the thermodynamic limit, the regular random graph can be identified with the Bethe lattice with proper boundary conditions
, because the local structure is tree-like (figure~\ref{fig:lattice}) and the typical length of the loop on the lattice diverges as $\log N$, where $N$ is number of sites. In this situation we can apply the cavity method. To fix the notation as well as for completeness of presentation, we present a short review of the cavity method for the generalized Ising model on a locally tree-like  lattice.

The effective field at site $i$ is calculated iteratively by a partial trace from $c-1$ spins connected to the current site. The independent effects of $c-1$ interactions have been passed to the current site $i$ in terms of the cavity field $h_i$ and cavity biases $\{u_j\}$, 
the definitions of which are given by
\begin{eqnarray}
h_i  &=& \sigma _i H + \frac{1}{\beta }\sum_{j=1}^{c - 1} {\tanh ^{ - 1} \left( {\tanh \left( {\beta J_{ij_1 \cdots j_{p-1}} } \right) \prod\limits_{k=1}^{p-1} {\tanh \left( {\beta h_{j_k} } \right)} } \right)} \\
&=& \sigma_i H + \sum_{j=1}^{c-1} {u_j(J_{ij_1 \cdots j_{p-1}},h_{j_k}) }, 
\label{recursion1}
\end{eqnarray}
and
\begin{equation}
\beta u_j  = \tanh ^{ - 1} \left( {\tanh \left( {\beta J_{ij_1 \cdots j_{p-1}} } \right) \prod\limits_{k=1}^{p-1} {\tanh \left( {\beta h_{j_k} } \right)} } \right),
\label{biasrecursion}
\end{equation}
where $\beta$ is the inverse temperature, $j$ labels $c-1$ interactions and $k$ labels $p-1$ spins per interaction on the layer previous to the current site $i$ (figure \ref{fig:lattice}).

After very many steps of these update rules, the effective field is obtained as
\begin{equation}\label{cfcent}
h^{(\rm{c})} = \sigma _i H + \sum_{j=1}^{c} {u_j(J_{ij_1 \cdots j_{p-1}},h_{j_k})  },\label{hc}
\end{equation}
and the marginal probability that site $i$ takes $S_i = \pm 1$ is
\begin{equation}
\chi^{S_i} = \frac{e^{\beta h^{(\rm{c})} S_i}}{2\cosh \beta h^{(\rm{c})}}.
\end{equation}
The Bethe free energy $f$ is calculated from the sum of the site and the bond terms \cite{Katsura:79,Nakanishi:81,Bowman:82} as
\begin{eqnarray}
\nonumber -\beta f &=& -\left( c - 1 \right) \log \left[ 2 \cosh \beta h^{(\rm{c})} \right] \\
&+& \frac{c}{p}\log \left[ \Tr\limits_{S_i} \exp\left( \beta J_{i_1 \cdots i_{p}} S_{i_1} S_{i_2} \cdots S_{i_{p}} + \textstyle{\sum_i^{p} h_i S_i} \right) \right].
\label{freeenergy}
\end{eqnarray}

If we neglect the possibility of RSB for the regular random graph, the distribution of the cavity field at the $l$th step $P^l\left( h \right) $ is updated as~\cite{M'ezard:09}
\begin{equation}
P^{l+1} \left( {h} \right) =  {\int { \left[ \delta (h - \sigma _i H - \sum\limits_{j = 1}^{c - 1} {u_j\left( J_{ij_1 \cdots j_{p-1}}, h_{j_k} \right) } ) \right]_{J,H} \prod\limits_{j = 1}^{c - 1}\prod\limits_{k = 1}^{p - 1} {P^{l} \left( {h_{j_k} } \right)dh_{j_k} }  }},
\label{cavityit}
\end{equation}
where $\left[ \cdots \right]_{J,H}$ denotes the average over the random interactions $J_{ij_1 \cdots j_{p-1}}$ and fields $\sigma _i$. 
If the distribution of the cavity field converges, the $P \left( h \right) = \lim_{l \rightarrow \infty} P^{l} \left( h \right)$ satisfies the following self-consistent equation,
\begin{equation}
P \left( {h} \right) =  {\int { \left[ \delta (h - \sigma _i H - \sum\limits_{j = 1}^{c - 1} {u_j\left( J_{ij_1 \cdots j_{p-1}}, h_{j_k} \right) } ) \right]_{J,H} \prod\limits_{j = 1}^{c - 1}\prod\limits_{k = 1}^{p - 1} {P \left( {h_{j_k} } \right)dh_{j_k} }  }},
\label{cavitysc}\end{equation}
and the distribution of the effective field is derived from
\begin{equation}
P^{(\rm{c})} \left( {h} \right) =  {\int { \left[ \delta (h - \sigma _i H - \sum\limits_{j = 1}^{c} {u_j\left( J_{ij_1 \cdots j_{p-1}}, h_{j_k} \right) } ) \right]_{J,H} \prod\limits_{j = 1}^{c}\prod\limits_{k = 1}^{p - 1} {P \left( {h_{j_k} } \right)dh_{j_k} }  }}.
\label{cavitycentral}
\end{equation}

Sometimes the RS solution is not correct and we need an RSB ansatz, which is much more complicated even for the one-step RSB (1RSB) solution. In order to draw phase diagrams of the present model, we use the 1RSB cavity method with the Parisi parameter $m=1$ (to be distinguished from magnetization)~\cite{Montanari:08, Zdeborov'a:09}. To this end, we define the message $\psi^{S_i}\left( h_i \right) = e^{\beta h_i S_i}/2\cosh \beta h_i$ which denotes the marginal probability of $S_i = \pm 1$ in the absence of the spin of the next step. The joint probability distribution of messages of the RS and 1RSB ($m=1$) ansatz is updated as~\cite{Montanari:08, Zdeborov'a:09}
\begin{eqnarray}
\nonumber P_{S_i}\left( \psi^{S'_i}_{\rm{1RSB}}, \psi^{S'_i}_{\rm{RS}} \right) &=& \sum_{ \{ S_{j_k }\} } \int{ \left[{ \delta \left( \psi^{S'_i}_{\rm{1RSB}} - F\left( \{ \psi^{S'_{j_k}}_{\rm{1RSB}} \} \right) \right) \delta \left( \psi^{S'_i}_{\rm{RS}}- F\left(\{ \psi^{S'_{j_k}}_{\rm{RS}} \} \right) \right) }\right. } \\
\nonumber  && \cdot \Bigl.{ w\left( \{ S_{j_k} \} | S_i\right) }  \Bigr]_{J,H} \prod\limits_{j = 1}^{c - 1}\prod\limits_{k = 1}^{p - 1} {P_{S_{j_k}}\left( \psi^{S'_{j_k}}_{\rm{1RSB}}, \psi^{S'_{j_k}}_{\rm{RS}} \right) d\psi^{S'_{j_k}}_{\rm{1RSB}} d\psi^{S'_{j_k}}_{\rm{RS}} },
\end{eqnarray}
where 
\begin{eqnarray}
w\left( \{ S_{j_k} \} | S_i\right) =  \frac{  \exp\left( \beta S_i \sum_{j} J_{ij_{1} \cdots j_{p-1} } \prod_{k} S_{j_k} \right)   \prod_{j}\prod_{k} \psi^{S'_{j_k}=S_{j_k}}_{\rm{RS}}}{\sum_{ \{ S_{j_k} \} } \exp\left( \beta S_i \sum_{j} J_{ij_{1} \cdots j_{p-1} } \prod_{k} S_{j_k} \right) \prod_{j}\prod_{k} \psi^{S'_{j_k}=S_{j_k}}_{\rm{RS}} }, \\
  F\left( \{ \psi^{S_{j_k}} \} \right) = \frac{ \sum_{ \{ S_{j_k} \} } \exp\left( \beta S_i \sum_{j} J_{ij_{1} \cdots j_{p-1} } \prod_{k} S_{j_k} + \beta H \sigma_i S_i \right)   \prod_{j}\prod_{k} \psi^{S_{j_k}}}{ \sum_{ S_i } \sum_{ \{ S_{j_k} \} } \exp\left( \beta S_i \sum_{j} J_{ij_{1} \cdots j_{p-1} } \prod_{k} S_{j_k} + \beta H \sigma_i S_i \right) \prod_{j}\prod_{k} \psi^{S_{j_k}} }. \nonumber 
\end{eqnarray}
The initial condition for distribution $P_{S_i}$ is
\begin{equation}
     P^{\rm init}_{S_i}\left( \psi^{S'_i}_{\rm{1RSB}}, \psi^{S'_i}_{\rm{RS}} \right) =  P\left(\psi^{S'_i}_{\rm{RS}} \right) \delta\left( \psi^{S'_i}_{\rm{1RSB}} - \delta_{S'_i,S_i}  \right)
\end{equation}

In order to identify the 1RSB spin glass phase boundary, we define the complexity function $\Sigma$ as the difference between the free energies calculated with $\psi_{\rm{RS}}$ and $\psi_{\rm{1RSB}}$,
\begin{equation}
\Sigma = -\beta f_{\rm{RS}} + \beta f_{\rm{1RSB} \left( \textit{m}=1 \right) },
\label{complexity}
\end{equation}
where $f_{\rm{RS}}$ is obtained from equation~(\ref{freeenergy}). On the other hand, $f_{\rm{1RSB} \left( \textit{m}=1 \right) }$ is calculated from~\cite{Montanari:08, Zdeborov'a:09}
\begin{equation}
-\beta f_{\rm{1RSB} \left( \textit{m}=1 \right) } = -\beta f_{\rm{1RSB}  }^{\rm{site}} -\beta f_{\rm{1RSB} }^{\rm{link}} ,
\label{freeenergyrsb}
\end{equation}
where, by using the probability distribution of the marginal $P_{S_i}^{\left( \rm{c}\right)}\left( \chi^{S'_i}_{\rm{1RSB}}, \chi^{S'_i}_{\rm{RS}} \right)$,
\begin{eqnarray}
-\beta f_{\rm{1RSB}  }^{\rm{site}} &=& -\left( c - 1 \right) \sum_{S_i}\int{ \chi_{\rm{RS}}^{S'_i=S_i} \log \left[ 2 \cosh \beta h^{(\rm{c})}_{S_i} \right] {P_{S_i}^{\left( \rm{c}\right)}\left( \chi^{S'_i}_{\rm{1RSB}}, \chi^{S'_i}_{\rm{RS}} \right) d\chi^{S'_i}_{\rm{1RSB}} d \chi^{S'_i}_{\rm{RS}} } } \nonumber \\
\\
-\beta f_{\rm{1RSB}  }^{\rm{link}} &=& \frac{c}{p}\sum_{ \{ S_{i}\} } \int{ \Biggl[ \log \Bigl[ \Tr\limits_{S''_i} \exp\bigl( \beta J_{i_1 \cdots i_{p}} S''_{i_1} S''_{i_2} \cdots S''_{i_{p}} + \textstyle{ \beta \sum_i^{p} h_{S_i} S''_i } \bigr) \Bigr]} w'\left( \{ S_{i}\} \right) \Biggr]_J \nonumber \\
&& \cdot\prod\limits_{i = 1}^{p} {P_{S_i}\left( \psi^{S'_i}_{\rm{1RSB}}, \psi^{S'_i}_{\rm{RS}} \right) d\psi^{S'_i}_{\rm{1RSB}} d\psi^{S'_i}_{\rm{RS}} }
\end{eqnarray}
\begin{eqnarray}
\beta h^{(\rm{c})}_{S_i} = S'_i \tanh^{-1} \left( 2 \chi^{S'_i}_{\rm{1RSB}} - 1  \right), \;\;\; \beta h_{S_i} = S'_i \tanh^{-1} \left( 2 \psi^{S'_i}_{\rm{1RSB}} - 1  \right) \\
w'\left( \{ S_{i}\} \right) =  \frac{  \exp\left( \beta J_{i_1 \cdots i_{p}} S_{i_1} S_{i_2} \cdots S_{i_{p}} + \beta \sum_i^{p} h_{S_i} S_i \right)   \prod_{i} \psi^{S'_{i}=S_{i}}_{\rm{RS}}}{\sum_{ \{ S_{i} \} } \exp\left( \beta J_{i_1 \cdots i_{p}} S_{i_1} S_{i_2} \cdots S_{i_{p}} + \beta \sum_i^{p} h_{S_i} S_i \right)   \prod_{i} \psi^{S'_{i}=S_{i}}_{\rm{RS}} }.
\end{eqnarray}

Actual calculations to identify the phase boundaries are described in the next subsection by using the cavity and effective field distributions.

\subsection{Numerical implementation}

The cavity method is implemented numerically in terms of the population method which realizes the probability distributions, $P^{(\rm{c})}\left( h \right)$, $P \left( u \right)$ and $\{ P_{S_i} ( \psi^{S'_i}_{\rm{1RSB}}, \psi^{S'_i}_{\rm{RS}} ) \}$ by a large amount of representative samples. The number of variables representing the distribution we use is $N_\mathrm{pop}=10^6$. 

There is an ambiguity in the initial condition of the RS iteration equation. As typical initial conditions, we have used the following two conditions for the RS cavity method.
\begin{enumerate}
 \item Ferromagnetic: all surface spins being in the up-state, which corresponds to $u_j^\mathrm{init} = J_{ij_1 \cdots j_{p-1}}$.
 \item Free: all surface spins being indeterminate as $u_j^\mathrm{init} = \pm \epsilon$ where $\epsilon\to 0$. 
This initial condition induces the paramagnetic solution of equation~(\ref{recursion1}).
\end{enumerate}
These conditions possibly derive different convergent distributions. An example is a first-order phase transition. In this case, we should choose an appropriate initial condition for the iterative solution of equation~(\ref{cavitysc}) to find the correct solution having the lowest free energy among all solutions of equation~(\ref{cavitysc}). Then the other solutions are metastable states due to the nature of the first-order phase transition.

We have performed at least $10,000$ updates of the cavity iteration until the distributions converge. After that, we have measured the physical values from the average of additional $20,000$ iterations.
 
Physical values characterizing the ferromagnetic and spin-glass phases are numerically calculated using the population method. The disorder averaged spontaneous magnetization is given by 
\begin{equation}
m = - \lim_{H_0 \rightarrow 0}\left[ \Part{f}{H_0}{} \right]_{J,H}=\int {d h P^{(\rm{c})}\left(h \right)\tanh \left( \beta h \right) },
\label{magdef}
\end{equation}
where $H_0$ is an additional uniform external field. The staggered magnetization is also calculated as
\begin{equation}
\hat{m} = -\left[ \Part{f}{H}{} \right]_{J,H}=\int {d h \left( { P_{+}^{(\rm{c})}\left(h \right) - P_{-}^{(\rm{c})}\left(h \right) } \right) \tanh \left( \beta h \right) },
\label{staggeredm}
\end{equation}
where $P^{(\rm{c})}_{\pm}$ denotes the distributions of the effective field having positive and negative signs of external field, respectively. The distribution of the cavity field satisfies
\begin{equation}
P^{(\rm{c})} \left( {h} \right) = P_{+}^{(\rm{c})} \left( {h} \right) + P_{-}^{(\rm{c})} \left( {h} \right).
\label{phnormalize}
\end{equation}

The onset of a spin-glass phase based on the local stability of the RS ansatz is signaled by the divergence of the spin-glass susceptibility defined as
\begin{equation}
\chi_\mathrm{SG} = \sum_G{ \left[ \left( \Part{\left< S \right>_0}{h_G}{} \right)^2 \right]_{J,H} }=\sum_G{ \left[ \left( \Part{\left< S \right>_0}{u_0}{} \Part{u_G}{h_G}{} \right)^2 \left( \Part{u_0}{u_G}{}  \right)^2 \right]_{J,H} },
\end{equation}
where the subscript $G$ denotes the index from a fixed site $0$ of the graph to site $G$ along the unique path. The divergence of $\chi_\mathrm{SG}$ is caused when
\begin{equation}
\lim_{G \rightarrow \infty} \left(c-1\right)^G \left[ \left( \Part{\left< S \right>_0}{u_0}{} \Part{u_G}{h_G}{} \right)^2 \left( \Part{u_0}{u_G}{}  \right)^2 \right]_{J,H} \ne 0.
\end{equation}
The essential part of the divergence is $\left[ \left( \partial{u_0}/\partial{u_G}  \right)^2 \right]_{J,H}$. This is implemented as an effect of a small perturbation to the cavity bias in numerical calculations~\cite{Pagnani:03,Krzakala:05,Jorg:08,Krzakala:10,Matsuda:10}. The numerical evaluation of the factor $\partial u_{0}/\partial u_G$ is 
straightforward,
\begin{equation}
\Part{u_0}{u_G}{} \approx
\frac{{u_0 \left( {u_G  + \Delta u_G } \right) - u_0
 \left( {u_G } \right)}} {\Delta u_G },
\label{eq:ATnum}
\end{equation}
where $\Delta u_G$ is a small perturbation which we set to $10^{-4}$. We prepare two replicas of identical convergent population of $\left\{ u_i \right\}$ and introduce a uniform perturbation into only one of the two replicas. Then we observe the square average of the variation, $(1/N_{\rm pop})\sum_{i=1}^{N_{\rm pop}}(u_{i}(u_G  + \Delta u_G) - u_{i}(u_G))^2$, after $10,000$ updates of the iteration equation~(\ref{biasrecursion}) with the same bond and site randomness for the two replicas.

The 1RSB spin-glass phase is signaled by the complexity function defined in equation~(\ref{complexity}). With temperature decreasing, the complexity function $\Sigma$ would jump from zero to a positive value at the dynamical 1RSB transition temperature and become zero at the equilibrium 1RSB transition temperature~\cite{Montanari:08, Zdeborov'a:09}. We have calculated this value for three-body ($p = 3$) systems as shown in the next section.

\section{Phase diagrams}
\label{numericalsection}
The results of the numerical implementation described in the previous section are shown here in terms of the phase diagrams under various conditions.

\subsection{The random-field Ising model with two-body purely ferromagnetic interactions}

Firstly we show our investigation for the purely ferromagnetic random-field Ising model ($\rho=1$) with two-body interactions ($p=2$) and connectivities $c=3$ and $4$. The absence of a spin-glass phase has been proved for this case ($\rho=1$, $p=2$) irrespective of the connectivity~\cite{Krzakala:10}.

At zero temperature, the spinodal point of the ferromagnetic phase (i.e. the largest $H$ admitting a ferromagnetic fixed point) has been solved as $H_\mathrm{F}^\mathrm{RF} = \lim_{M\rightarrow \infty} \left( \left( c-2 \right) + 2/M \right) = c-2$, where $M$ is interpreted as the number of spins in ferromagnetic clusters at zero temperature~\cite{Bruinsma:84}. A ferromagnetic phase transition is observed when the size of the cluster $M$ is infinity, but also there are Griffiths singularities at $H = \left( c-2 \right) + 2/M $ for $M=1,2,\cdots$ only at zero temperature. For finite temperatures the phase boundary has been evaluated approximately by several techniques~\cite{Bruinsma:84,Wohlman:84,Hartzstein:85,Galam:85,Swift:94,Bleher:98,Nowotny:01,Krzakala:10},
and with the cavity method in \cite{Krzakala:10}. Here we give more details for finite-temperature case. 

\begin{figure}[tb]
\begin{minipage}[t]{0.5\hsize}
\includegraphics[width=1.\linewidth]{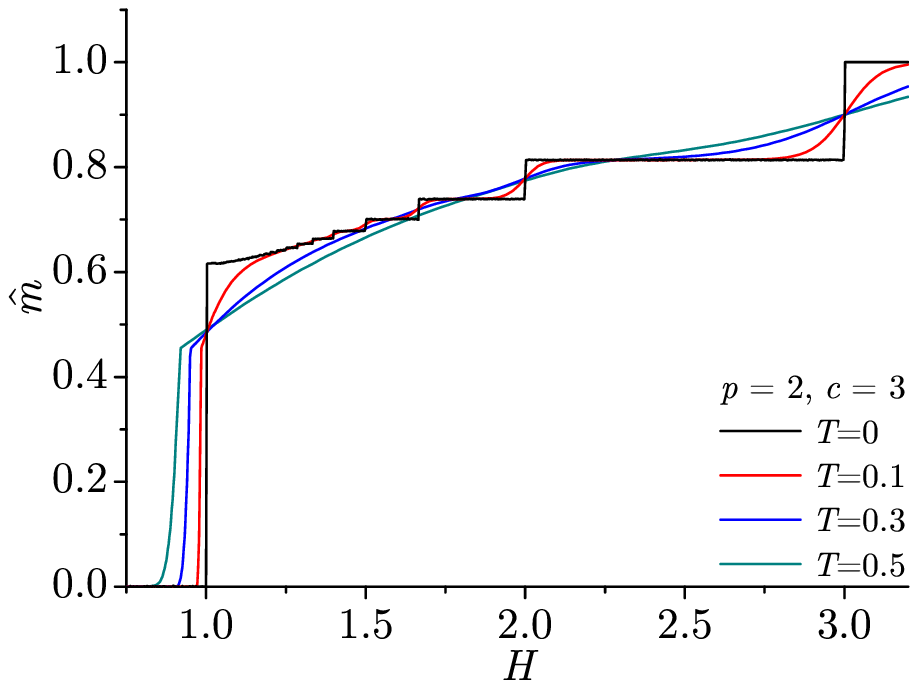}
\end{minipage}
\begin{minipage}[t]{0.5\hsize}
\includegraphics[width=1.\linewidth]{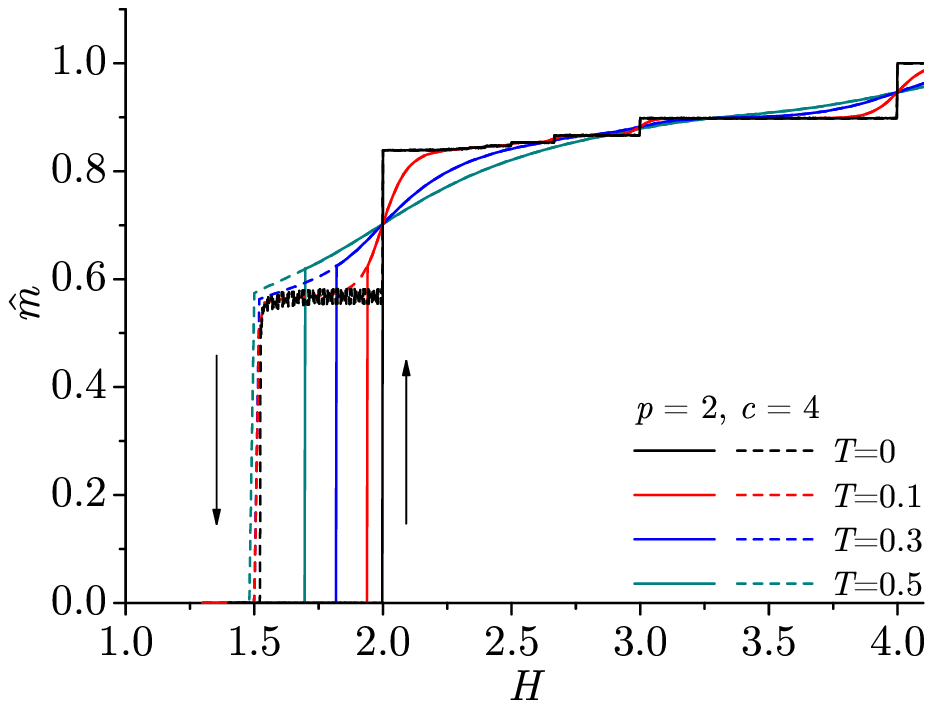}
\end{minipage}
\caption{
(Colour online) Staggered magnetization for the random field Ising model ($\rho=1$) on the regular random graph with two-body interactions ($p=2$) and connectivity $c=3$ (left) and $c=4$ (right). Solid and dashed lines denote the numerical results with ferromagnetically fixed and free initial conditions, respectively. The system has singularities at $H = \left( c-2 \right) + 2/M $ ($M=1,2,\cdots$) at zero temperature, and the spinodal point of the ferromagnetic phase is given by $H_\mathrm{F}^\mathrm{RF} = c-2$.

\label{fig:mk2c34}}
\end{figure}

When $p=2$ and $ c\le3$, the transition is second order. Otherwise it is first order where hysteresis is observed in the magnetization~\cite{Bruinsma:84,Wohlman:84,Rosinberg:09}. Figure \ref{fig:mk2c34} shows the staggered magnetization for $p=2, \, c=3$ and $4$ from $T=0$ to $0.5$. Only at zero temperature, the staggered magnetization jumps at $H = \left( c-2 \right) + 2/M$ for both cases of $c=3$ and $4$ but does not for finite temperature except at transition points.

For $c=4$, the convergent distribution $P^{(\rm{c})}\left( h \right)$ has two solutions. The magnetization exhibits different results depending on the initial condition for equation~(\ref{cavitysc}) at low temperature, while there is a unique solution for $c=3$. Thus, for the systems with $c=4$, the transition is first order at low temperature. The higher transition temperature represents the limit of stability of the ferromagnetic solution and the lower is for the paramagnetic solution. The two transition points may be identified by  changing the field $H$, i.e. increasing or decreasing $H$ as indicated in arrows in figure~\ref{fig:mk2c34}.
\begin{figure}[tb]
\begin{minipage}[t]{0.5\hsize}
\includegraphics[width=1.\linewidth]{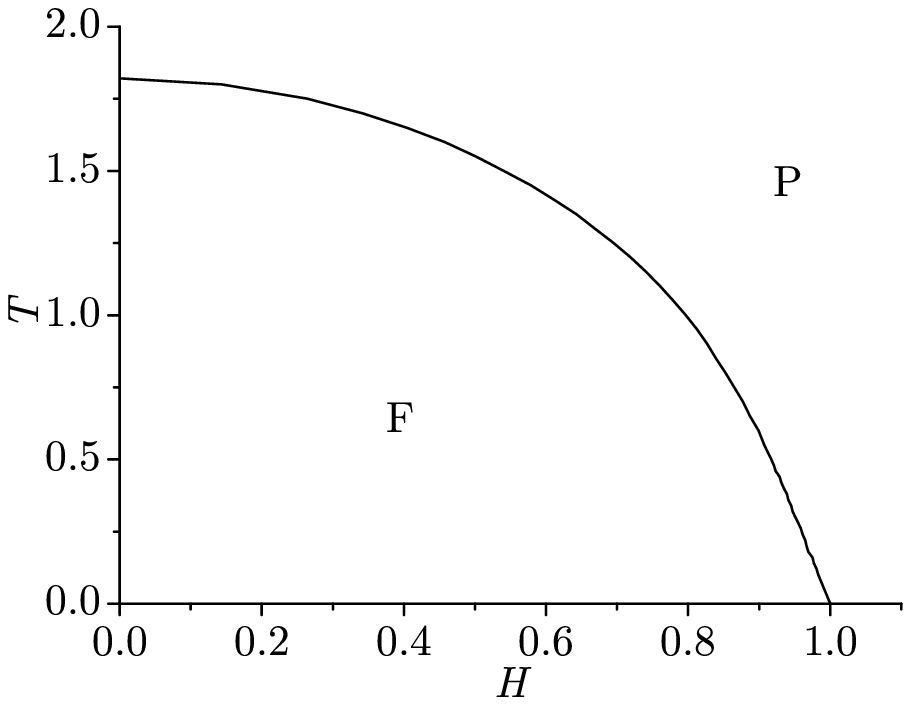}
\end{minipage}
\begin{minipage}[t]{0.5\hsize}
\includegraphics[width=1.\linewidth]{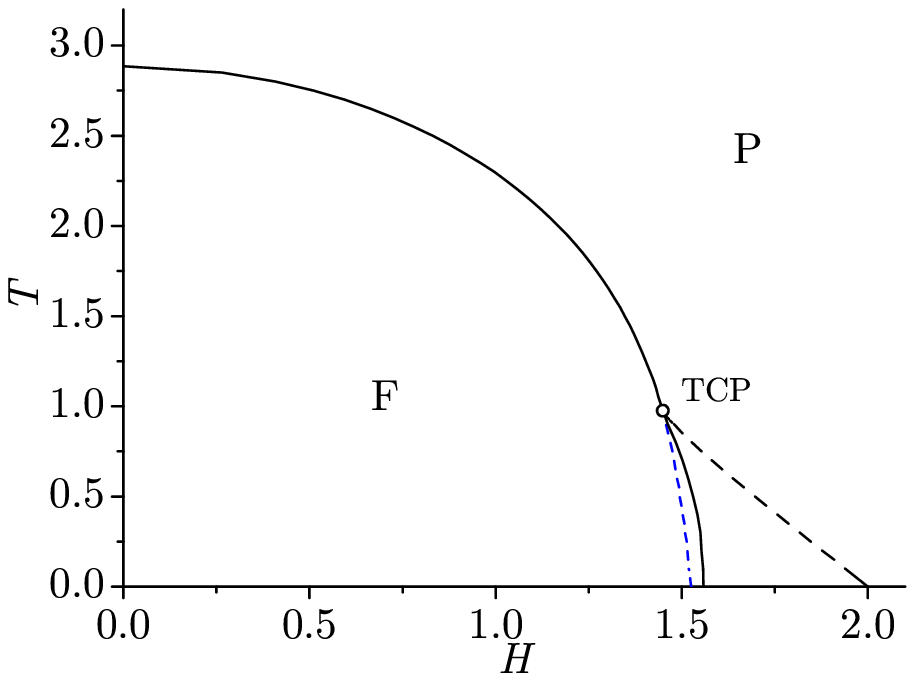}
\end{minipage}
\caption{
(Colour online) Phase diagrams of the random field Ising model with $p=2$ on the regular random graph with $c=3$ (left) and $c=4$ (right). The symbols stand for the paramagnetic (P) and ferromagnetic (F) phases, and $\rm{TCP}$ is for the tricritical point. Zero field transition points are given by $T_\mathrm{F}^\mathrm{RF}\left( H=0 \right) = 1/\tanh^{-1}\left( 1/\left( c-1 \right) \right) \approx 1.820$ ($c=3$) and $2.885$ ($c=4$) for the ferromagnetic phase. In the right panel, spinodal lines are drawn  dashed for paramagnetic (blue) and ferromagnetic (black) phases.
\label{fig:pdk2c34}}
\end{figure}

The equilibrium phase boundary lies between these two critical-field values, which correspond to the metastability limits of respective phases. As shown on the right panel of figure~\ref{fig:pdk2c34}, these two metastability limits merge at a tricritical point, and a second-order phase transition takes over beyond the tricritical point for lower $H$ and higher $T$. The equilibrium phase boundary is calculated as the points where free energies for the two solutions have the same value by using equation~(\ref{freeenergy}).

\subsection{The random-field $\pm J$ Ising model with two-body interactions}

No spin-glass phase is observed under the two-body pure-ferromagnetic condition as shown in the previous subsection but it readily appears by the introduction of a small amount of bond randomness. In fact, the zero-temperature para-ferro transition point of the pure ferromagnetic case, as shown in figure~\ref{fig:pdk2c34}, turns out to be a multicritical point where paramagnetic, ferromagnetic and spin-glass phases coexist as will be shown below.

Figure~\ref{fig:pdk2c34t0} shows the zero temperature ($T=0$) phase diagram of the $\pm J$ Ising model in random fields with $p=2$ and $c=3$ (left) and $c=4$ (center and right). We see that the transition point at $\rho=1$ and $T=0$ behaves as a multicritical point.
\begin{figure}[tb]
\includegraphics[width=1.\hsize]{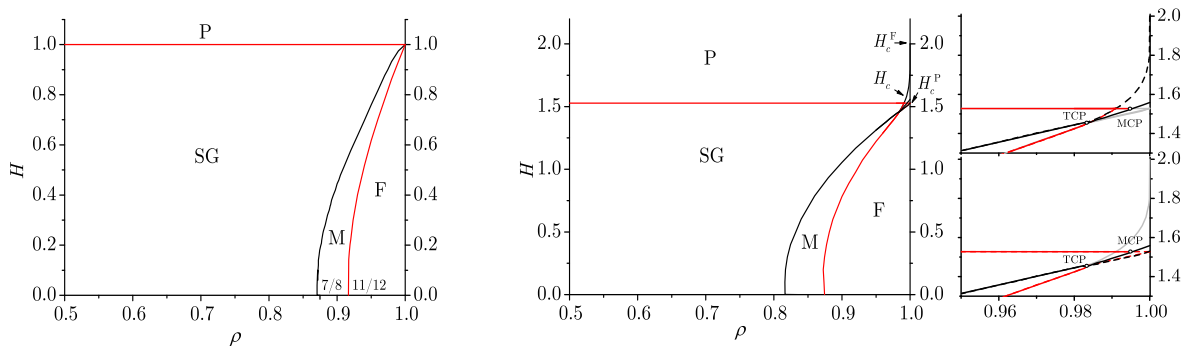}
\caption{
(Colour online) Zero temperature ($T=0$) phase diagram of the $\pm J$ Ising model in random fields with two-body interactions with connectivity $c=3$ (left panel) and $c=4$ (center and right two panels). The red lines denote the boundaries of the spin glass phase. The  symbols stand for the spin glass (SG) and mixed (M) phases and the tricritical (TCP) and multicritical (MCP) points. The mixed phase is the region surrounded by the spin glass and ferromagnetic phase boundaries. 
For $c=3$ the zero field critical probability is analytically known: one has $\rho_c^\mathrm{SG} = 11/12$ for the spin-glass phase~\cite{Kwon:88,Castellani:05} while $\rho_c^\mathrm{F}({\rm{RS}})=7/8$ at the RS level~\cite{Kwon:88,Castellani:05} and $\rho_c^\mathrm{F}({\rm{1RSB}})\approx0.857$ at the 1RSB level.
In the limit of $H \rightarrow 0$ and $T \rightarrow 0$, our value of $\rho_c^\mathrm{F}$ is a bit lower than the RS one of references \cite{Kwon:88,Castellani:05} since we work with real cavity fields whereas references \cite{Kwon:88,Castellani:05} have integer ones. It is still, however, larger than the 1RSB value which is usually very close from the exact value.
For $c=4$ (center and right panels) the phase diagram has a phase coexistence region. The right panel shows a blow up of the center panel in the vicinity of $\rho=1$ with the ferromagnetically-fixed initial condition (i) in Section 2.2 (upper) and the free initial condition (ii) (lower). The equilibrium phase boundaries are shown by solid lines and spinodal lines are by dashed lines. The spin glass-paramagnetic boundaries at $H=1.0$ ($c=3$) and $H=1.5275(5)$ ($c=4$) show no dependence on $\rho$.

\label{fig:pdk2c34t0}}
\end{figure}
This means that the introduction of any small amount of bond randomness produces a spin-glass phase  as an equilibrium state ($c=3$) and a metastable state ($c=4$). The spin-glass has ferromagnetically ordered state in the mixed phase.

For $c>3$, a first-order ferromagnetic transition has been found at $\rho=1$. The first-order transition line extends from $\rho=1$ to some $\rho_\mathrm{TCP}<1$, and a locally stable ferromagnetic solution exists in this area. The phase diagram (right panel of figure~\ref{fig:pdk2c34t0}) is thus more complex than the case $c=3$ due to the nature of the first-order phase transition.

\begin{figure}[tb]
\begin{minipage}[t]{0.5\hsize}
\includegraphics[width=1.\linewidth]{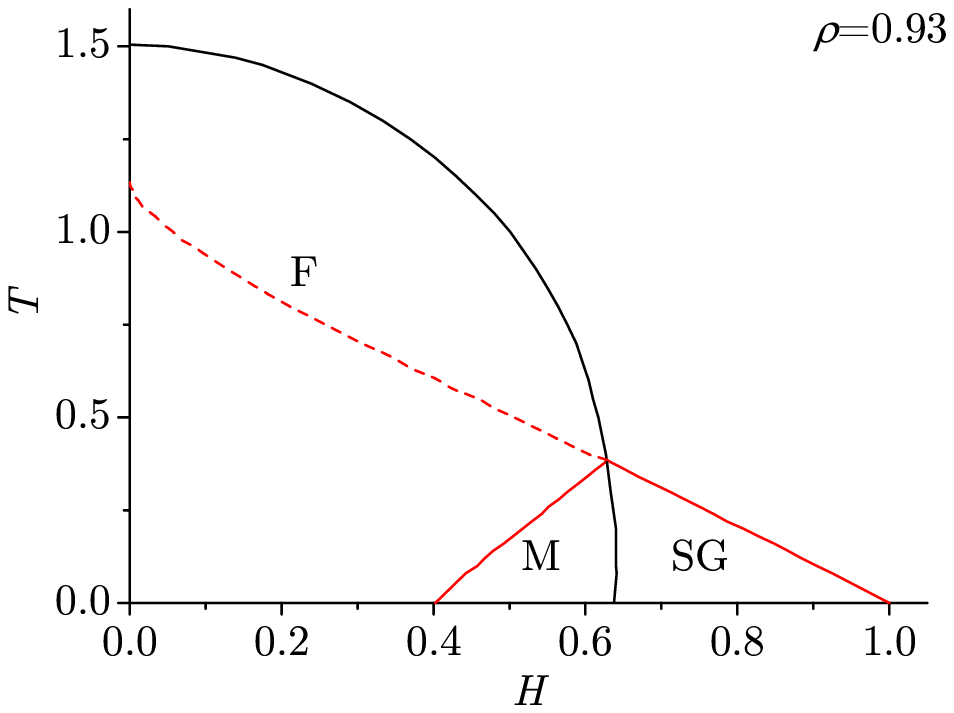}
\end{minipage}
\begin{minipage}[t]{0.5\hsize}
\includegraphics[width=1.\linewidth]{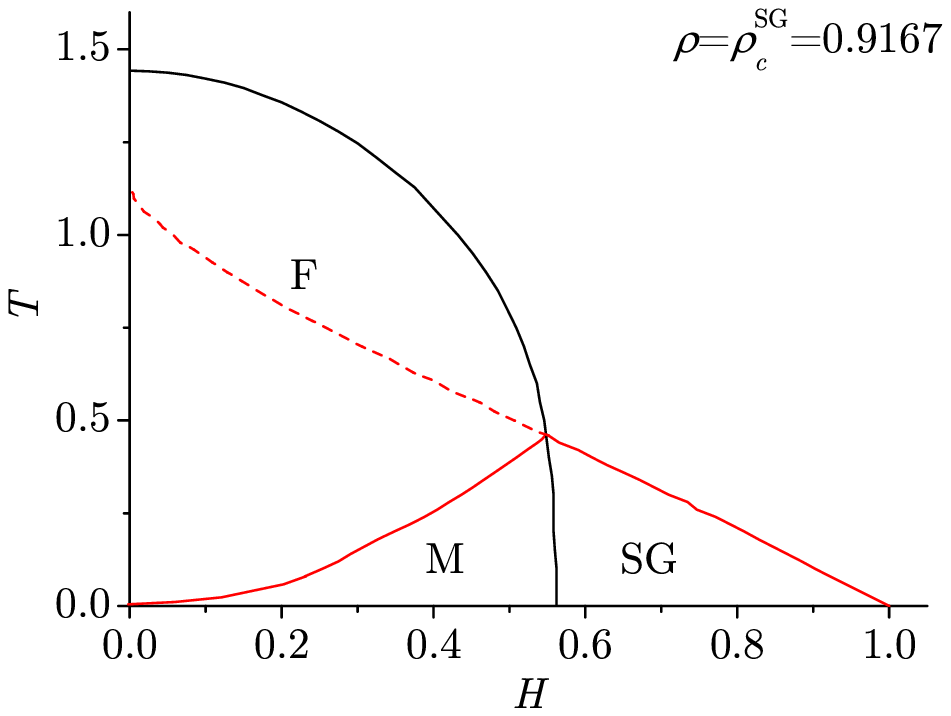}
\end{minipage}
\begin{minipage}[t]{0.5\hsize}
\includegraphics[width=1.\linewidth]{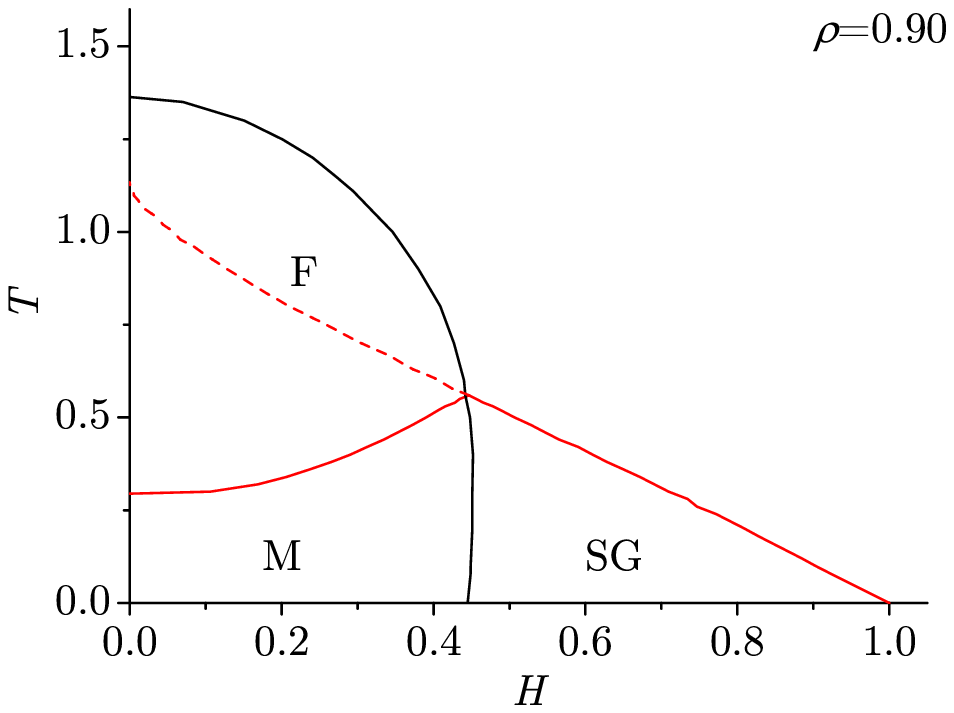}
\end{minipage}
\begin{minipage}[b]{0.5\hsize}
\includegraphics[width=1.\linewidth]{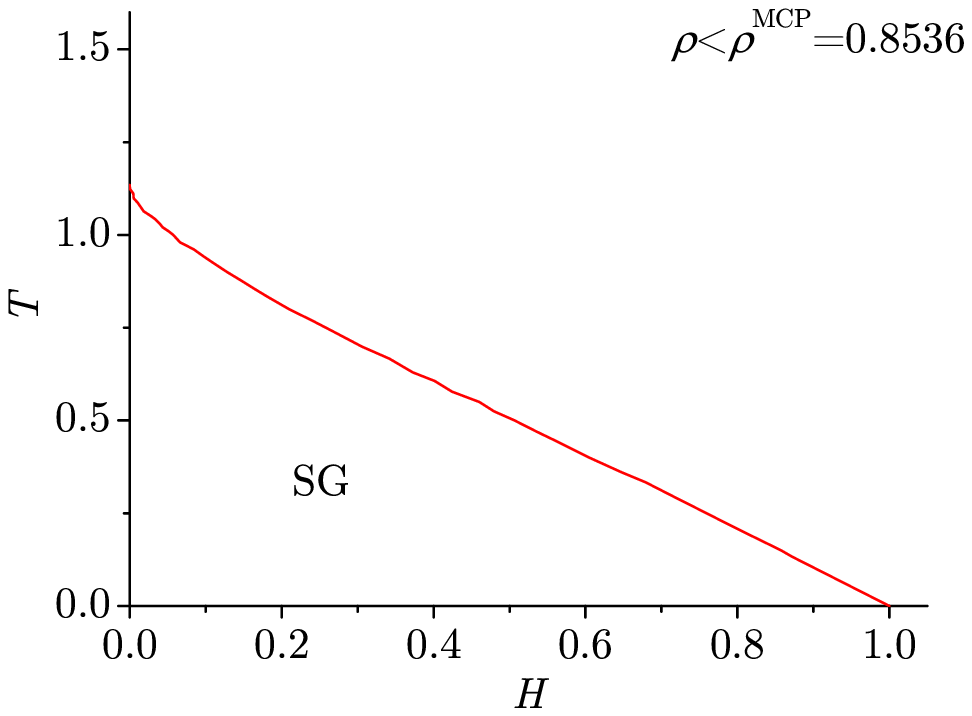}
\end{minipage}
\caption{
(Colour online) Phase diagrams of the two-body $\pm J$ Ising model in random fields with $c=3$. The ferromagnetic phase shrinks as $\rho$ is decreased and the spin-glass phase appears from behind the ferromagnetic phase. The dashed line stands for the spin-glass phase boundary for $\rho=0.5$ drawn for the guide of the eyes. It does not represent a real phase boundary.
\label{fig:pdk2c3ft}} 
\end{figure}

The phase boundary between the spin glass and paramagnetic phases is independent of the ferromagnetic bias $\rho$.  This behaviour is clearly shown in the finite temperature phase diagram: Figure~\ref{fig:pdk2c3ft} depicts phase diagrams for $\rho<\rho^\mathrm{MCP}=\left( 2 + \sqrt{2} \right) /4(=0.8536)$ 
($\rho^\mathrm{MCP}$ here is defined such that at this density $T_c^F=T_c^{SG}$) 
and $\rho=0.90,\, \rho_c^\mathrm{SG}(=11/12=0.9167)$ and $0.93$ with $c=3$. The ferromagnetic phase vanishes for $\rho<\rho^\mathrm{MCP}$, and $ \rho_c^\mathrm{SG} = 11/12$ is defined as the limit below which ($\rho<\rho_c^\mathrm{SG}$) the $H=0$ system has a finite temperature spin glass phase. For $\rho<\rho^\mathrm{MCP}$, the spin-glass phase boundary shows no dependence on $\rho$ down to $1/2$. On the other hand, the ferromagnetic phase exists for $\rho>\rho^\mathrm{MCP}$, and phase boundaries of ferromagnetic and mixed phases depend on $\rho$, which shrinks with decreasing $\rho$.

\subsection{Three-body interactions}

The independence of $\rho$ of the spin glass-paramagnetic boundary appears also in systems with three-body interactions in random fields with and without randomness in interactions.

\begin{figure}[tb]
\includegraphics[width=1.\hsize]{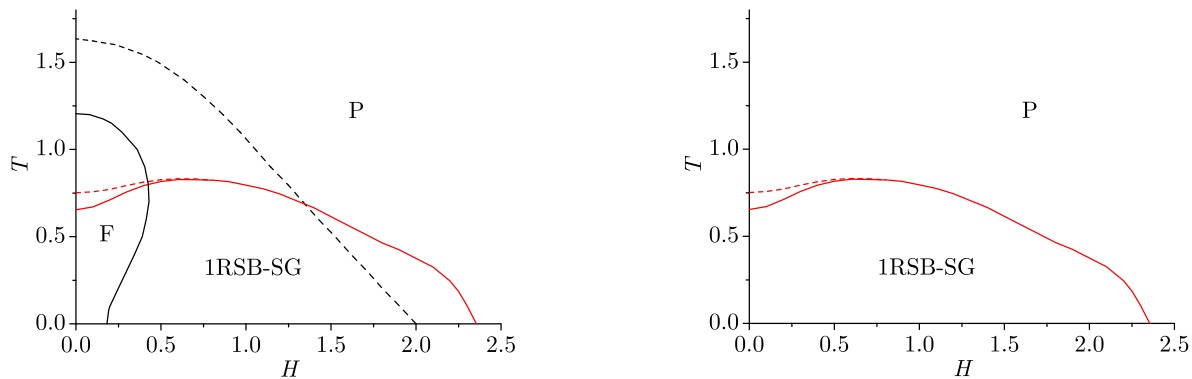}
\caption{
(Colour online) Phase diagrams for the three-body random-field Ising model ($c=4$) for $\rho=1$ (left) and $\rho=1/2$ (right).
The black dashed line denotes the spinodal line of the ferromagnetic phase. Solid black line is the equilibrium phase boundary of the ferromagnet. The red dashed line is the dynamical glass transition and the solid red line the static (Kauzmann) glass transition. The spin-glass phase exists for any $\rho$ unlike the two-body interacting system (figure~\ref{fig:pdk2c34}), but there is no mixed phase. The spin-glass phase boundaries in two panels coincide. The phase diagram shown in the right panel does not depend on the ferromagnetic bias $\rho$ as long as we consider only spin glass and paramagnetic solutions.

\label{fig:pdk3c4p1}}
\end{figure}

The left panel of figure~\ref{fig:pdk3c4p1} is the phase diagram obtained from the 1RSB cavity method for $\rho=1$. 
The dashed black line denotes the spinodal of ferromagnetic phase. The red dashed line is the dynamical spin glass temperature, and the solid red line is the static (Kauzmann) spin glass temperature. The solid black line is the equilibrium phase boundary obtained from the comparison of ferromagnetic and paramagnetic free energies.
Note that the lower part of the ferromagnetic-spin glass phase boundary is derived from a comparison between free energies of the ferromagnetic and spin-glass solutions based on the RS ansatz. Thus this phase boundary is inaccurate. However, the spin glass phase surely exists adjacent to the ferromagnetic phase in the high-$H$, low-$T$ region. Therefore the proof of the absence of a spin glass phase in the random-field Ising model does not apply beyond its limit of two-body interactions~\cite{Krzakala:10}. 

The right panel of figure~\ref{fig:pdk3c4p1} is the phase diagram for $\rho=1/2$. The spin glass phase boundaries (red, solid and dashed lines) in the left and right panels are identical to each other. In addition, the phase diagram in the right panel is obtained also for arbitrary $\rho$ for magnetization fixed to zero. The $\rho$ dependence does not appear outside the ferromagnetic phase for the three-body system. It is expected that the ferromagnetic phase boundary again shrinks with decreasing $\rho$ and the spin glass phase hidden behind the ferromagnetic phase gradually emerges.

The spin glass phase for the three-body system is observed as 1RSB. By analogy with the study for fully-connected graphs~\cite{Gillin:01}, the full-RSB (FRSB) spin glass phase may exist in the high-$H$ region. In order to draw the boundary between 1RSB and FRSB phases, however, we need to analyze the stability of the 1RSB solution, which is beyond the scope of this paper.

At zero field, we obtained dynamical and static/Kauzmann 1RSB spin-glass transition temperature as $T_d^{\rm{1RSB}} = 0.7526(1)$ and $T_c^{\rm{1RSB}} = 0.6547(3)$, respectively, and the complexity function $\Sigma \left(T = T_d^{\rm{1RSB}} \right) = 0.05415(3)$ by $2.0\times 10^5$ updates of the population with $10^6$ variables. These values agree well with those in references~\cite{Franz:01,Montanari:04}.

\subsection{The $\pm J$ Ising model without field}

\begin{figure}[tb]
\begin{minipage}[t]{0.5\hsize}
\includegraphics[width=1.\linewidth]{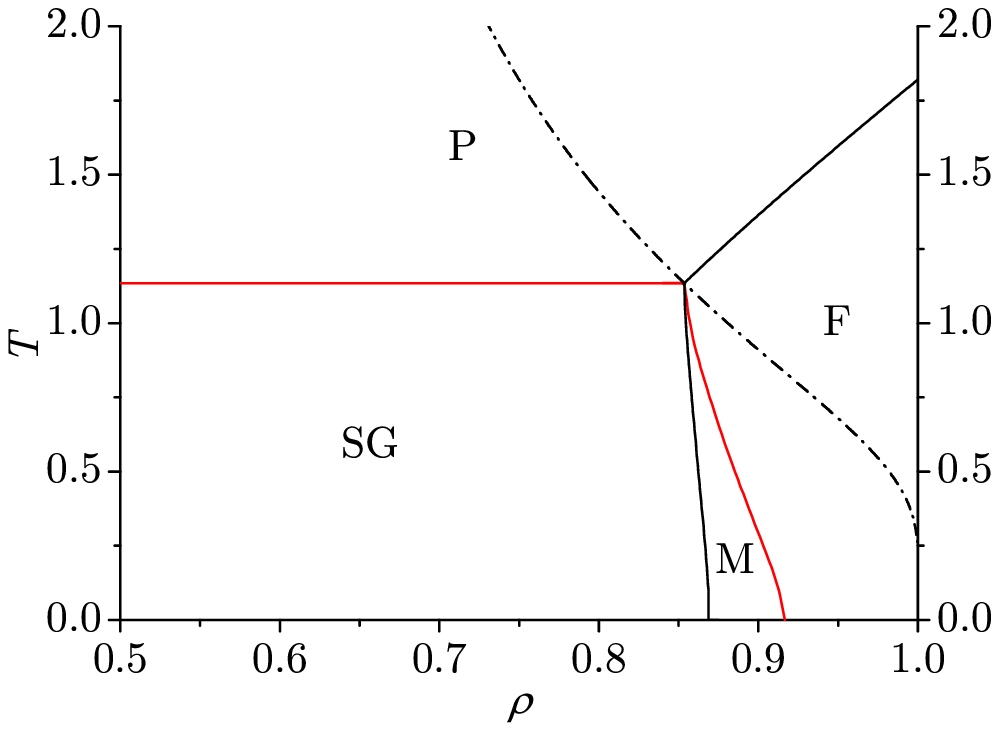}
\end{minipage}
\begin{minipage}[t]{0.5\hsize}
\includegraphics[width=1.\linewidth]{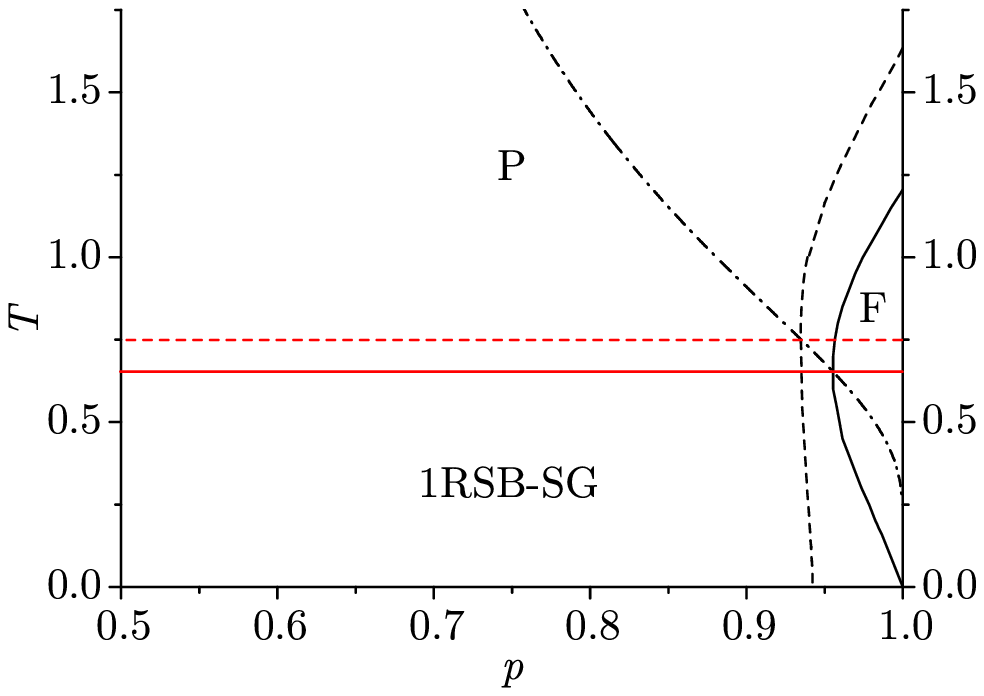}
\end{minipage}
\caption{
(Colour online) Phase diagrams of the $\pm J$ model without external field for $p=2,\,c=3$ (left) and $p=3,\,c=4$ (right). The solid and dashed lines show the static/Kauzmann and dynamical transition points, respectively, and the chained black lines denote the Nishimori line defined as $e^{2\beta} =\rho/\left( 1-\rho\right)$~\cite{Nishimori:81}. In the right panel, the phase boundary between ferromagnetic and spin glass phases below the equilibrium multicritical point is obtained from the comparison between the the ferromagnetic and RS paramagnetic solutions. 
This boundary is thus inaccurate and at the 1RSB level these are instead {\it almost} straight lines below the multicritical point in both cases \cite{Castellani:05,Krzakala:10b}.
\label{fig:pdk2c3pT}}
\end{figure}

Also for the system without external field (the usual $\pm J$ Ising model), we observe the independence of $\rho$ of the phase boundary between spin glass and paramagnetic phases as shown in figure~\ref{fig:pdk2c3pT} (see also figure 1 in \cite{Krzakala:10b}).
The spin glass-para phase boundary is horizontal for both static and dynamical critical temperature lines. The transition points for $p=2$ and $c=3$, the case of the left panel, does not depend on $\rho$ and are analytically given by $T_\mathrm{F}\left( H=0 \right) = 1/\tanh^{-1}\left( 1/\left( c-1 \right) \right)$ for the ferromagnetic phase and $T_\mathrm{SG}\left( H=0 \right) = 1/\tanh^{-1}\left( 1/\sqrt{ c-1 } \right)$ for the spin-glass phase~\cite{Thouless:86,Carlson:90,Carlson:90a,Kabashima:03}. 
The spin glass phase appears even in the pure ferromagnetic system. This feature is seen also in figure~\ref{fig:pdk3c4p1}.

Note that the phase boundary between ferromagnetic and spin glass phases below the multicritical point in the right panel 
is not exact, because it is obtained from the free energies based on the RS ansatz, i.e. comparison between the ferromagnetic and paramagnetic solutions. One needs to investigate the free energy in the 1RSB spin glass phase at finite temperature in order to find the correct phase boundary, as e.g. in \cite{Castellani:05,Krzakala:10b}.

\section{Distribution of zeros on the complex-field plane}
\label{zeros}

Distribution of partition function zeros is one of the unique viewpoints of the phase diagram~\cite{Yang:52, Lee:52, Nishimori:11}, and it can also be understood as the generalized (complex valued) phase diagram. In this section, we confirm the $\rho$-independence of system properties outside the ferromagnetic phase, as found in the previous section, on the complex-field plane. The distribution of zeros clearly shows the difference and similarity between the system for $\rho=1$ and $0.5$. All results of distributions of zeros have been obtained based on the RS ansatz.

In reference~\cite{Matsuda:10}, the authors have found the relationship between the distribution of complex cavity field $P_{H,\beta}^{(\rm{c})}\left(h \right)$ and the two-dimensional (real) density of zeros $g_2 \left( H \right) $ as
\begin{equation}
g_2 (\hat H\, ) = P_{H,\beta}^{(\rm{c})}\left(h=\pi \mathrm{i} /2 \beta \right),
\end{equation}
where $\hat H= 2 \beta H$, and $h$ and $\beta$ are complex values. 
Also, the one-dimensional density of zeros in general is calculated as~\cite{Lee:52},
\begin{equation}
2 \pi g_1 (\hat H\, )= \frac{1}{2} \Re \left[ \hat m (  \hat H + \epsilon ) - \hat m ( \hat H - \epsilon ) \right],
\end{equation}
where $\epsilon$ is an infinitesimally small complex number which is normal to the curve of the one-dimensional density of zeros and $\hat m = - \partial f / \partial H$ is also of complex value. If the external field is uniform, $\hat m$ is the usual magnetization whereas $\hat m$ is defined as the staggered magnetization shown in equation~(\ref{staggeredm}) for systems with random fields. These relations enable us to see the distribution of zeros in the infinite system size.

Transition points are identified as the locations where zeros touch the real axis on the complex-parameter plane, because the  free energy has a singularity at the location of partition function zeros. The one-dimensional distribution of zeros crosses the real axis at the ferromagnetic transition point whereas the spin glass phase is characterized that the continuous (two-dimensional) distribution of singularities touches the axes of real field and temperature~\cite{Matsuda:10}.

\begin{figure}[tb]
\begin{minipage}[t]{0.5\hsize}
\includegraphics[width=1.\linewidth]{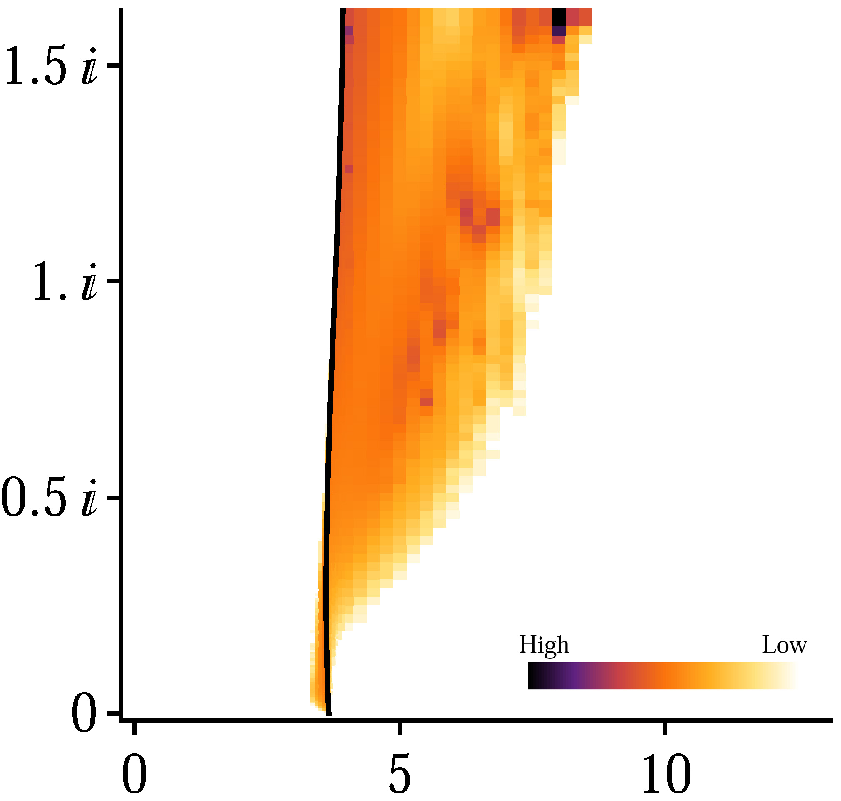}
\end{minipage}
\begin{minipage}[t]{0.5\hsize}
\includegraphics[width=1.\linewidth]{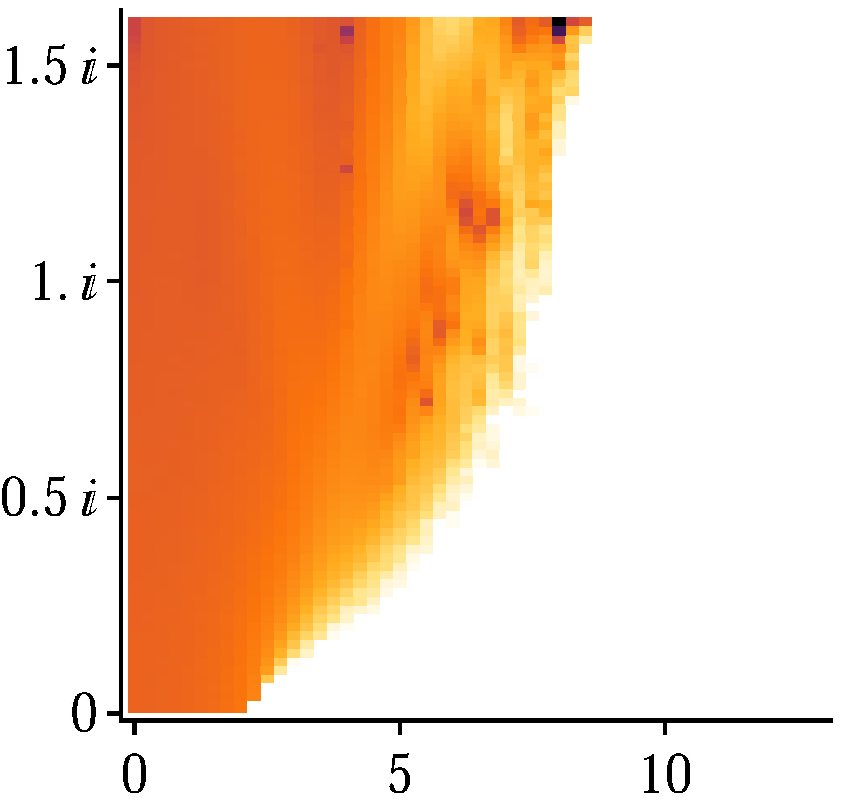}
\end{minipage}
\caption{
(Colour online) The distributions of partition function zeros on the complex $2 \beta H$ plane for the two-body $\pm J$ Ising model with random fields for $\rho=1$ (left) and $\rho=0.5$ (right) at $T=0.5$. The one-dimensional distribution of zeros (drawn in a thick line) runs vertically in the left panel. The distribution to the right of this one-dimensional distribution is the same as the distribution on the right panel in the same region, showing its independence of $\rho$.
\label{fig:dzk2c3}}
\end{figure}

In figure~\ref{fig:dzk2c3}, we show the distribution of zeros on the complex-field plane for the two-body $\pm J$ Ising model with random fields for $\rho=1$ (left panel) and $\rho=0.5$ (right panel) with connectivity $c=3$ at $T=0.5$. For numerical evaluations of the distribution of zeros, the complex-$\hat H$ plane has been split into cells by dividing the real axis from $\Re \hat H = 0$ to $12$ with an increment $0.25$ and the imaginary axis from $\Im \hat H = 0.02$ to $\pi /2$ with an increment of $0.02$. The distribution of zeros has a symmetry with respect to the real and imaginary axes, and therefore only the first quadrant is shown. Both $g_1$ and $g_2$ are plotted in the same figure and coloured in a logarithmic scale; a black dot shows a very high density.

In the left panel of figure~\ref{fig:dzk2c3}, the ferromagnetic phase (defined as the region to the left of the one-dimensional distribution) is bounded by the thick black line representing $g_1 > 0$ where $\Re \, \hat m$ jumps. It crosses the real axis at the (real) ferromagnetic transition point. The region to the left side of $g_1$ may thus be characterized as the ferromagnetic phase. Compared to the right panel, the distribution of zeros outside the ferromagnetic phase is the same as the one for $\rho=0.5$. Therefore the distribution of zeros outside the ferromagnetic phase does not depend on $\rho$ while the phase boundary of the ferromagnetic phase does. The two-dimensional distribution of zeros $g_2$ to the left side of $g_1$ barely exists in the vicinity of $g_1$. It would expand and lead to the mixed phase with decreasing $\rho$ as shown in figure~\ref{fig:pdk2c3ft}.

\begin{figure}[t]
\begin{minipage}[t]{0.5\hsize}
\includegraphics[width=1.\linewidth]{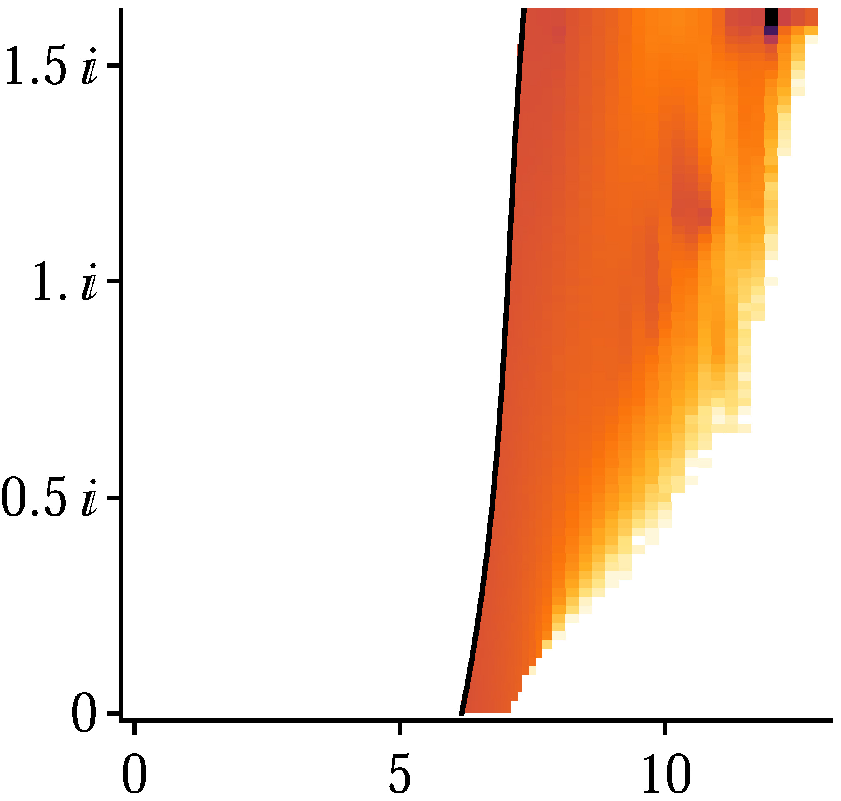}
\end{minipage}
\begin{minipage}[t]{0.5\hsize}
\includegraphics[width=1.\linewidth]{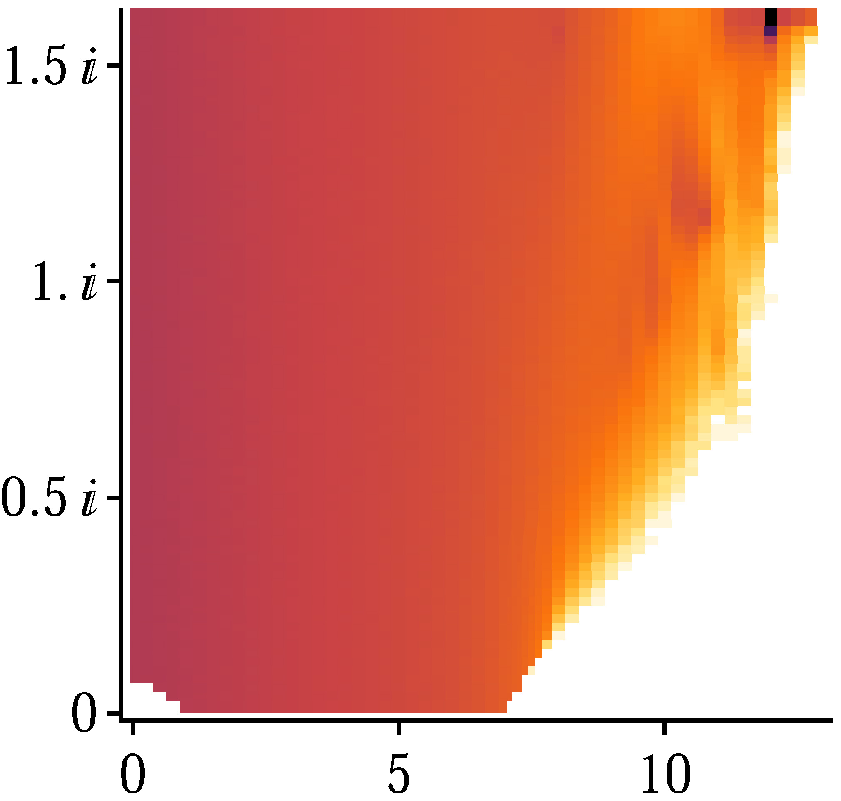}
\end{minipage}
\caption{
(Colour online) The distributions of partition function zeros on the complex $2 \beta H$ plane for $\rho=1$ (left) and $\rho=0.5$ (right) for three-body interactions with the ferromagnetically-fixed initial condition at $T=0.5$. There is the spin glass phase even for the pure ferromagnetic system, which is shown as the continuous singularity of the system in the left panel. The right panel can be obtained also from the pure ferromagnetic random-field Ising model with the non-magnetized solution of equation~(\ref{cavitysc}).
\label{fig:dzk3c4}}
\end{figure}
For the three-body system, on the other hand, there exists the spin glass phase for the random-field Ising model even when $\rho=1$. The left panel of figure~\ref{fig:dzk3c4} shows the distribution of zeros for the system of $\rho=1$ with the ferromagnetic solution of equation~(\ref{cavitysc}). The spin glass phase is characterized by the continuous distribution of singularities, see the region to the right side of $g_1$ along the real axis in the left panel of figure~\ref{fig:dzk3c4}. Comparison to the distribution for the case of $\rho=0.5$ (right) suggests the independence of $\rho$ outside the ferromagnetic phase for the three-body system as well. Note that the distribution of zeros shown in the right panel of figure~\ref{fig:dzk3c4} is also obtained from the case of $\rho=1$ with the non-magnetized solution of equation~(\ref{cavitysc}). Due to the absence of spontaneous magnetization originated from a first-order phase transition, the system property is completely independence of $\rho$ for some solutions. This is an important difference between the two- and three-body systems. Incidentally, there is no two-dimensional zeros in the ferromagnetic phase of the left panel of figure~\ref{fig:dzk3c4}, which is consistent with the absence of a mixed phase in the three-body random field Ising model (the left panel of figure~\ref{fig:pdk3c4p1}).

\section{Gauge transformation}
\label{gauge}

In previous sections, we have investigated the phase diagrams of the two- and three-body $\pm J$ Ising model with and without random fields. Our numerical results suggest that many of the important properties of the system do not depend on the ferromagnetic bias $\rho$. 
This behavior can be explained using a gauge transformation, and this appeared for various special cases of models on random graphs in the literature, see e.g.~\cite{Krzakala:08, Zdeborov'a:10}. Since we do not know of an article with a detailed and general explanation of the associated reasoning, we provide one in this section. We split the argument in two parts. Firstly we show the exact $\rho$-independence for trees (i.e. cycle-free graphs) and then for the thermodynamic limit for random graphs (i.e. locally tree-like graphs). Finally we discuss limitations of this $\rho$-independence.     

We recall that, on a cycle-free graph with free boundary conditions, the thermodynamics properties of the $\pm J$ model with symmetric distribution of random fields do not depend on the ferromagnetic bias $\rho$. To show this, consider a realization of disorder with $\rho$, and another realization with $\rho=1$ (the pure ferromagnet) and the following gauge transformation
\begin{eqnarray}
      S_i  &\to& \tau_i S_i \\
      J_{ij_1 \cdots j_{p-1}}  &\to& J_{ij_1 \cdots j_{p-1}} \tau_i \tau_{j_1} \cdots \tau_{j_p-1}, \label{gauge1}
\end{eqnarray}
where $\tau_i = \pm 1$. As the graph is cycle-free, choose a node at random, call it a root, then proceed from the root to the leaves with the following transformation. Start with $\tau_i=1$ for all $i$. Consider all interactions descendent from the root for which $J=-1$. For each of those, choose one descendant node $j$ (arbitrarily), set $\tau_j=-1$ and recalculate all the $J$'s around node $j$. Now consider the second generation and again for all interactions with $J=-1$ set $\tau_j=-1$ for one of the descendants $j$. Continue this iteratively until the leaves are reached. This transformation is uncorrelated with the way the external random field was chosen. Hence any system (of any size) with symmetric distribution of the random field and free boundary conditions can be directly mapped to the purely ferromagnetic case. 

The above mapping was possible for tree graphs with free boundary conditions because those do not contain any frustration. It is well known that the $\pm J$ models on lattices with loops and with frustration cannot be exactly mapped to the ferromagnetic (unfrustrated) case. This is simply because the above gauge transformation conserves the frustration. The $\pm J$ model without frustration is called the Mattis model~\cite{Mattis:76}, and it is well known that this one is equivalent to the ferromagnetic Ising model, which generalizes to the Mattis model in symmetric random fields. For models with frustration on random graphs, the reason for $\rho$-independence is more involved.

Consider now a random field $\pm J$ model with ferromagnetic bias $\rho = 1/2$. The iteration equation for the cavity bias for this model is
\begin{equation}
\beta u_{j\rightarrow i}  = \tanh ^{ - 1} \left( {\tanh \left( {\beta J_{ij_1 \cdots j_{p-1}} } \right) \prod\limits_{k=1}^{p-1} {\tanh \left( {\beta \sum_{l}^{c-1}{u_{l \rightarrow j_{k}}} + \beta \sigma_{j_k} H } \right)} } \right).
\label{cavsympmj}
\end{equation}
Now, if we consider the gauge transformation $J_{ij_1 \cdots j_{p-1}} = \tilde J_{ij_1 \cdots j_{p-1}} \tau_i \tau_{j_1} \cdots \tau_{j_p-1}$ where $\tilde J$ has an arbitrary ferromagnetic bias $\rho$ and symmetric distribution $P_\tau\left( \tau \right)$, equation~(\ref{cavsympmj}) changes to
\begin{equation}
\beta \tau_i u_{j\rightarrow i}  = \tanh ^{ - 1} \left( {\tanh \left( {\beta \tilde J_{ij_1 \cdots j_{p-1}}} \right) \prod\limits_{k=1}^{p-1} {\tanh \left( {\beta \sum_{l}^{c-1}{\tau_{j_k} u_{l \rightarrow j_{k}}} + \beta \tau_{j_k} \sigma_{j_k} H } \right)} } \right).
\end{equation}
Notice that the symmetric distribution of $\tau$ guarantees the symmetric distribution of $J_{ij_1 \cdots j_{p-1}} = \tilde J_{ij_1 \cdots j_{p-1}} \tau_i \tau_{j_1} \cdots \tau_{j_p-1}$. If we write $\tilde u_{j\rightarrow i} = \tau_i u_{j\rightarrow i}$, the above equation is rewritten as
\begin{equation}
\beta \tilde u_{j\rightarrow i}  = \tanh ^{ - 1} \left( {\tanh \left( {\beta \tilde J_{ij_1 \cdots j_{p-1}}} \right) \prod\limits_{k=1}^{p-1} {\tanh \left( {\beta \sum_{l}^{c-1}{ \tilde u_{l \rightarrow j_{k}}} + \beta \tau_{j_k} \sigma_{j_k} H } \right)} } \right).
\label{eq:aftergauge}
\end{equation}
Since $\tau$ and $\sigma$ are independent, the distribution $P_{\tau \sigma} (\tau \sigma)$ is also symmetric. Thus, this iteration equation~(\ref{eq:aftergauge}) can be interpreted as that for a model with an arbitrary $\rho$. However, note that the distribution of $\tilde u_{j}$ obtained through the gauge transformation is always symmetric, because $\tau$ is symmetric and independent of $u$. 
Since $P_{u} \left(u \right) $ for $\rho=1/2$ is symmetric, $P_{u} \left(u \right) = P_{\tilde u} \left( \tilde u \right) $ holds.
 Hence, the solution which has a symmetric distribution of cavity field does not depend on the ferromagnetic bias $\rho$.

In the argument for $\rho$-independence on random graphs in the previous paragraph we used two assumptions: (i) the cavity method (with proper RSB scheme if needed) provides an exact solution for the system under consideration. (ii) the distribution of cavity fields is symmetric around zero. Assumption (i) limits the $\rho$-independence to the thermodynamic limit of models on random tree-like graphs. 

We have observed the symmetry of cavity fields on random regular graphs anytime the magnetization of the resulting solution was zero
, i.e. for the paramagnet and the (non-mixed) spin glass phase. On the other hand, this explains why the ferromagnetic solution depends crucially on $\rho$. Moreover, the solution corresponding to the symmetric cavity fields formally continues to exist even in the ferromagnetic region. If the ferromagnetic transition is of a second order ($p=2$), this non-magnetized phase is locally unstable towards magnetized solutions. For a first order ferromagnetic phase transition ($p>2$), it is locally stable and hence is a physically observable metastable phase. This is shown in figure~\ref{fig:pdk3c4p1} and the right panel of figure~\ref{fig:dzk3c4}. In fact, in this case ($p>2$), it is not easy to find the ferromagnetic phase without the use of ferromagnetically biased initial conditions. For instance, simulated annealing~\cite{Kirkpatrick:83} does not work for the pure ferromagnetic three-body system without field, although the ground state is trivial. The three-spin ferromagnet is actually used as one of the hardest examples of optimization problems with a known solution \cite{Haanpaa:05}. 

Coming back to condition (ii) on the symmetry of the cavity fields, we want to stress that in general fixing the magnetization to zero does not ensure that the corresponding cavity field must be symmetric and $\rho$-independent. The reason is that the symmetric cavity field gives zero magnetization but the inverse is not always correct (see equation~(\ref{magdef})). A counter-example is given by the ferromagnetic model without field at magnetization fixed to zero on Erd\"{o}s-Renyi random graph which is not equivalent to the corresponding spin glass model \cite{Zdeborov'a:10}. 

Note now that for $\rho=1/2$ the random bimodal field model is equivalent to the uniform field spin glass model. Hence on the $T$-$H$ plane, the spin-glass phase boundary for the present model is the same as the de Almeida-Thouless line for the symmetric $\pm J$ Ising model in a uniform field when we equate the strength of random field $H$ with that of uniform field. Thus the spin-glass phase boundary outside the ferromagnetic phase lies on the de Almeida-Thouless line. This property does not depend on $\rho$, but the ferromagnetic phase boundary does. On the $H$-$\rho$ and $T$-$\rho$ planes as well, the spin-glass phase boundary is a horizontal line outside the ferromagnetic phase, because an arbitrary location on the phase diagram except for the ferromagnetic phase can be mapped onto the corresponding point for the symmetric $\pm J$ Ising model.

We also note that values of all gauge-invariant quantities do not depend on $\rho$ outside the ferromagnetic phase. For instance, the exact solution for the internal energy in the paramagnetic phase of the purely ferromagnetic system can be applied to the whole paramagnetic phase of the $\pm J$ model. It is consistent with the fact that the internal energy on the Nishimori line equals to that for the pure ferromagnetic system without field as $E= - N_B \tanh \beta$, where $N_B$ is the number of the interactions~\cite{Nishimori:81}. Furthermore, the values of gauge-invariant non-equilibrium quantities do not depend on $\rho$. For instance, the auto correlation function $\left[ \left<  S_i\left(t=0 \right) S_i\left(t=\tau \right) \right> \right]_J$ has this property.  This property, combined with the fact that on the Nishimori line the ferromagnetic configuration has properties of an equilibrium configuration, was recently used in \cite{Krzakala:10b} to argue about relationship between the glassy dynamics and melting above a first order phase transition.

\section{Conclusion}

We have investigated the $p$-body $\pm J$ Ising model with and without random fields on the regular random graph with fixed connectivity $c$. We have focused on the phase boundary of the spin glass phase and have revealed its essential independence of the ferromagnetic bias in the interactions.

In section~\ref{numericalsection} we have numerically evaluated the phase diagrams by using the cavity method. Our results indicate for arbitrary $p$, $\rho$ and $c$ that the spin-glass phase boundary outside the ferromagnetic phase coincides with the boundary at $\rho=1/2$. This property appears as the horizontal phase boundary of the spin-glass phase on the $T$-$\rho$ planes. Actually every non-magnetized solution of the self-consistent cavity equation of the distribution of the cavity field is completely independent of $\rho$. We have also shown the distribution of zeros on the complex field plane in section~\ref{zeros}. Based on the comparison between the systems for $\rho=1/2$ and $\rho=1$, the distributions are equal to each other except in the ferromagnetic phase. In section~\ref{gauge} we gave a theoretical explanation for the independence on the ferromagnetic bias using a gauge transformation and the fact that the cavity solution for random graph (with proper level of RSB) is expected to be exact.  

A very remarkable result is the presence of a static spin glass phase in the pure ferromagnet for the three-body interactions, see figure~\ref{fig:pdk3c4p1}. The absence of such a phase was proved for ferromagnetic two-body interactions \cite{Krzakala:10}. Here we showed that this proof cannot generalize to models with many-body interactions.

\section{Acknowledgements}
This work was partially supported by CREST, JST. YM is grateful for the financial support provided through the Japan Society for the Promotion of Science (JSPS) Research Fellowship for Young Scientists Program.

\section*{References}

\providecommand{\newblock}{}

\end{document}